\documentclass[apj]{emulateapj}   
\usepackage{amsmath}
\usepackage{natbib}
\usepackage{amssymb}
\usepackage{epsf}
\usepackage{epsfig}
\usepackage{graphics}     
\usepackage{graphicx}   
\usepackage{url}           
\usepackage{bm}           
\usepackage{epstopdf}

\newcommand{\mbf}{\mathbf}
\newcommand{\be}{\begin{equation}}
\newcommand{\ee}{\end{equation}}

\begin{document}

\title{Role of magnetic field strength and numerical resolution in simulations of the heat-flux driven buoyancy instability}
\author{Mark J. Avara\altaffilmark{1}, Christopher Reynolds\altaffilmark{1,2}, Tamara Bogdanovi\'c\altaffilmark{3}}
\altaffiltext{1}{Department of Astronomy, University of Maryland, College Park, MD 20740, USA; mavara@astro.umd.edu, chris@astro.umd.edu, tamarab@gatech.edu}
\altaffiltext{2}{Joint Space Science Institute (JSI), University of Maryland, College Park, MD 20740, USA}
\altaffiltext{3}{Center for Relativistic Astrophysics, School of Physics, Georgia Tech, Atlanta, GA 30332, USA}
\date{\today}

\begin{abstract}
	The role played by magnetic fields in the intracluster medium (ICM) of galaxy clusters is complex.  The weakly collisional nature of the ICM leads to thermal conduction that is channeled along field lines.   This anisotropic heat conduction profoundly changes the stability of the ICM atmosphere, with convective stabilities being driven by temperature gradients of either sign.  Here, we employ the Athena magnetohydrodynamic code to investigate the local non-linear behavior of the heat-flux driven buoyancy instability (HBI), relevant in the cores of cooling-core clusters where the temperature increases with radius.  We study a grid of 2-d simulations that span a large range of initial magnetic field strengths and numerical resolutions.  For very weak initial fields, we recover the previously known result that the HBI wraps the field in the horizontal direction thereby shutting off the heat flux.  However, we find that simulations which begin with intermediate initial field strengths have a qualitatively different behavior, forming HBI-stable filaments that resist field-line wrapping and enable sustained vertical conductive heat flux at a level of 10--25\% of the Spitzer value.  While astrophysical conclusions regarding the role of conduction in cooling cores require detailed global models, our local study proves that systems dominated by HBI do not necessarily quench the conductive heat flux.
\end{abstract}


\section{Introduction}

	The intracluster medium (ICM) is a hot, weakly-collisional plasma comprising the majority of the baryonic mass in galaxy clusters. It has gained increasing attention in the last few decades since it provides a rich environment for the study of feedback, cosmology, low-density magnetohydrodynamics (MHD), and magnetogenesis. One area of recent focus is the study of dynamics induced by anisotropic conduction and anisotropic viscosity. The main goal of these studies has been to develop a better understanding of the role of magnetic fields in moderating heat conduction in the ICM. 
	
	Anisotropic conduction results from the fact that the electron gyroradius is much smaller than its mean free path and renders the ICM atmosphere buoyantly unstable to temperature gradients of either sign, relative to the direction of gravity. The outer region of the ICM, where virialization supports a negative temperature gradient, falls subject to the magnetothermal instability \citep[MTI;][]{balbus2000}. On the other hand, in the inner region of cool-core clusters the plasma density is high enough for bremsstrahlung cooling to turn over the temperature gradient and the atmosphere is unstable to the heat-flux driven buoyancy instability \citep[HBI;][hereafter Q08]{quataert2008}.
	
	Previous studies of the local non-linear evolution of the HBI using numerical simulations \cite[]{P&Q2008,2011MNRAS.413.1295M} found that the instability acts to quench vertical heat conduction by wrapping magnetic field lines horizontally thereby insulating each temperature layer from one another. If this strong insulation effect occurs in real clusters, it renders thermal conduction irrelevant in consideration of the thermal energy balance of the cooling core.
	
	In the absence of significant conduction, one must invoke AGN feedback or another type of turbulent stirring to stave off catastrophic cooling, but any such mechanism would simultaneously reorient some field lines vertically and conduction would again drive turbulence via the HBI. Therefore any self-consistent model of the ICM needs to both realistically capture the physics of the HBI as well as feedback processes \cite[]{BR2008, Sharmaetal2009, Parrishetal2010, KunzSCBS2011}. 
	
	The HBI is fundamentally a weak-field instability, operating even when magnetic forces are completely negligible (provided that the field is strong enough to keep the electron gyroradius smaller than its mean free path).   For this reason, most previous studies of the HBI have assumed extremely weak initial fields, $\beta\gtrsim 10^{11}$ where $\beta\equiv P/P_{\rm mag}$ with $P$ being the thermal plasma pressure and $P_{\rm mag}$ the magnetic pressure.    However, the role of magnetic field/tension on the non-linear behavior of the HBI is poorly understood.   \citet[hereafter K2012]{KunzBogdanovicReynoldsStone2012} performed semi-global HBI runs with initial $\beta=10^5$ and found a qualitatively different behavior to that seen previously.  Instead of the field line wrapping and quenching of conduction seen by previous authors, K2012 found sustained conduction via magnetic filaments.  However, it was unclear whether the semi-global nature of the simulation or the stronger initial field was the main cause of this difference.   

This leaves open several questions.   Does magnetic tension play a defining role in shaping the non-linear behavior of the HBI?   Since tension forces are most relevant for highly curved field lines, is the numerical resolution typically used in simulations sufficient to fully capture the physics of the HBI?   What role do magnetic filaments play in the transport of heat in the ICM?
	
	In this paper we present new 2-d simulations in the local regime which cover the resolution-$\beta$ parameter space in an attempt to answer these questions. We explore how the non-linear dynamics of the weakly collisional ICM changes as the initial magnetic field strength is varied from complete convective stability (HBI completely suppressed by magnetic tension at the local scale) to the low field strength used in previous studies. In order to determine whether current simulations have sufficient grid resolution for all aspects of the physics of the HBI to be fully converged, we also study trends in fundamental measures of the non-linear state of the plasma as grid resolution is varied. 
	
	We principally find that, as suggested by inviscid simulations of K2012, the non-linear state of the HBI is qualitatively different than in previous studies when an intermediate (within a few orders of magnitude of HBI stability) initial magnetic field strength is chosen. In particular, there is enough magnetic flux to form sustained vertical filaments which prevent the quenching of vertical heat conduction in the ICM, saturating at a significant fraction of the Spitzer conductivity \cite[]{spitzer1962}. We present the results of our parameter space exploration and describe new insights into the physics of HBI growth as well as the formation of magnetic filaments, including a connection between the non-linear turbulent state of the plasma and the linear theory. Our results hold for moderate resolution 3-d simulations, validating the applicability of a 2-d study and the physical model we present. 
	
	The paper is organized as follows. In  \textsection\ref{sec:MHD} we briefly review the analytic description of the HBI focusing on aspects relevant for our results. In \textsection\ref{sec:numMeth} we discuss the numerical method, boundary conditions, and initial conditions. \textsection\ref{sec:Observations} presents the results of the simulations and describes major trends we observe, which are then placed into context of the physics of the HBI in \textsection\ref{sec:FormDyn}. In \textsection\ref{sec:3D} a complimentary 3-d simulation of moderate resolution is described, and \textsection\ref{sec:Discussion} provides a discussion of our work in the context of existing literature.

\section{MHD of Weakly Magnetized Plasmas}\label{sec:MHD}
	
	We use the standard MHD equations with a term added to the entropy equation for anisotropic thermal conduction along magnetic field lines. For easy comparison to recent literature we primarily follow the prescription for linear analysis laid out in \citet[hereafter BR10]{2010ApJ...720L..97B} while excluding the term for radiative cooling.
	
	We choose rationalized natural units with $k_B = c = \epsilon_{0} = \mu_{0} = 1 $ such that total energy density is given by 
	\begin{equation}  E = \epsilon+\rho\frac{\mathbf{v} \cdot \mathbf{v}}{2} + \frac{\mathbf{B} \cdot \mathbf{B} }{8\pi} - \frac{\rho \mbf{g}\cdot\mbf{x}}{(\gamma-1)}
	\end{equation}
	with the internal (i.e., thermal) energy density $\epsilon = P/(\gamma-1)$, where we assume the adiabatic index $\gamma = 5/3$ and the mean molecular weight $\mu=1$ (pure hydrogen, fully ionized, plasma). The third term on the right hand side is the magnetic energy density, or pressure, used in the definition of the plasma $\beta$-parameter 
	\be 
		\beta=\frac{8\pi P}{B^2}.
	\ee
	 The vector quantity $\mbf{x}=x\hat{x}+z\hat{z}$ defines position in the 2-d simulations. $\hat{z}$ points vertically upward and $\hat{x}$ points horizontally to the right. Then, the mass conservation, momentum conservation, induction, and entropy equations are, respectively,
	\begin{gather}
	\frac{\partial \rho}{\partial t} + \nabla \cdot (\rho \mathbf{v})=0,\\
	\rho \frac{D\mathbf{v}}{Dt} = \frac{(\nabla \times \mathbf{B})\times \mathbf{B} }{4\pi } - \nabla P + \rho\mathbf{g},\\
	\frac{ \partial \mathbf{B} }{\partial t} = \nabla \times(\mathbf{v}\times\mathbf{B}),\\
	\frac{D\ln P\rho^{-\gamma}}{Dt} = -\frac{\gamma-1}{P}\nabla\cdot\mathbf{Q} \label{eqn:entropy}
	\end{gather} 
	where $\rho$ is the mass density, $\mathbf{v}$ is the fluid velocity, $\mathbf{B}$ is the magnetic field vector, $P$ is the gas pressure, $\mathbf{g}$ is the gravitational acceleration, and $\mathbf{Q}$ is the conductive heat flux. $D/Dt\equiv\partial/\partial t+\mathbf{v}\cdot\nabla$ is the Lagrangian derivative.
	
	We define the anisotropic conductive heat flux $\mbf{Q}$, with $\mathbf{b}=\mathbf{B}/B$ the unit vector in the direction of the magnetic field, as 
	\be  \mathbf{Q}=-\chi\mathbf{b}(\mathbf{b}\cdot\nabla)T, \ee
	where T is the kinetic gas temperature and
	\[   \chi \simeq 6\times10^{-7}T^{5/2} \mbox{ erg cm$^{-1}$ s$^{-1}$ K$^{-1}$} \]
	is the thermal conductivity \cite[]{spitzer1962} in physical units. As in Q08 and BR10 we define $\kappa\equiv\chi T/P$ to be the anisotropic thermal diffusion coefficient. In natural units, $\kappa\approx10^{-2}$ is the Spitzer diffusion coefficient for a plasma of temperature 1 keV.  
	
	We ignore the ion component of the heat flux since it is smaller than the electron contribution by a factor of $(m_{i}/m_{e})^{1/2}\approx42$. \cite{KunzSCBS2011} showed that, despite this large ratio, viscosity can have a significant effect on the HBI for conductivity at the Spitzer value, as chosen here, but that this effect is severely diminished deep in cluster cores where the plasma is most collisional. For this work, we choose not to include viscosity so that we may probe the connection between the formation of magnetic filaments and the conduction driven physics of the HBI alone. Regardless, the presence of anisotropic viscosity seems to only enhance the formation of filaments \cite[]{2012MNRAS.422..704P}.

	\subsection{Background Equilibrium Conditions}\label{subsec:bec}
	We consider a vertically stratified atmosphere in which the gas is not self-gravitating and the extent of the domain is small enough to be in the local regime. We then approximate the gravitational acceleration to be a constant function of position $\mathbf{g}(z)=-g_{0}\hat{z}$. 
		
	The initial (unperturbed) magnetic field is purely vertical for all simulations presented. The simulations start with equilibrium initial conditions in which magnetic pressure is negligible compared to the thermal pressure. Moreover, a uniform initial field provides zero magnetic pressure gradient so field strength does not affect the equilibrium condition at all. Therefore the equilibrium state of the atmosphere is that of hydrostatic balance with
	\begin{equation}
	\frac{dP}{dz}=\mathbf{g}(z)\rho = -g_0 \rho.
	\end{equation}	 
	
	Given an initial temperature profile T(z), in order to show distinction between that of the HBI and that of the classical adiabatic (Schwarzschild) instability, we employ profiles that have entropy increasing with height, i.e., that are Schwarzschild stable. Thus, as will be shown in \textsection \ref{sec:numMeth}, we assume a simple power law representation for equilibrium temperature and density such that $\partial s/\partial z > 0$ where the entropy s is
	\[  s = \frac{P}{\rho^\gamma}. \]

	\subsection{Stability analysis}\label{subsec:wkb}
	We refer the reader to BR10 for a full construction of the Wentzel-Kramers-Brillouin (WKB) perturbation analysis using plane wave disturbances of the form exp($\sigma t + i\mathbf{k\cdot x}$) where $\mathbf{x}$ is the position vector, $\sigma$ is the formal growth rate, and the wavenumber $\mathbf{k}$ has Cartesian components. However, there are a few points in the analysis which are particularly relevant to the results in this paper.
	
	Without cooling, the linear analysis in BR10 results in the following dispersion relation:
	\begin{align} \label{eqn:disprel}
	\left(  \sigma + C   \right)(\sigma^2 + &(\mbf{k\cdot v_A})^2) \\
	&+ \frac{\sigma k_\bot^2 N^2}{k^2}+CK\frac{g}{k^2}\frac{\partial \ln T}{\partial z} = 0,
	\end{align}
	where 
	\be 
	C\equiv \left( \frac{\gamma-1}{\gamma}\right) \kappa (\mbf{k\cdot b})^2  ,
	\ee
	\be
	K\equiv (1-2b_z^2)k_\bot^2 + 2b_x b_z k_x k_z,
	\ee
	and the Brunt-Vaisala frequency, N, given by
	\be
	N^2=-\frac{1}{\gamma\rho}\frac{\partial P}{\partial z} \frac{\partial \ln s}{\partial z},
	\ee
	describes the buoyant response of an adiabatic plasma. $k_\bot$ is the component of $\mbf{k}$ perpendicular to the magnetic field direction. The square of the Alfvenic frequency is $\omega_A^2 \equiv (\mbf{k}\cdot \mbf{v_A})^2$.     
	
	This cubic dispersion relation encodes three conditions for stability, one of which encodes the HBI and MTI criterion of \cite{balbus2000} and Q08,
	\be \label{eqn:HBI_MTIcondition}
	K\frac{g}{k^2}\frac{\partial\ln T}{\partial z} + \omega_A^2 > 0,
	\ee
	where it is convenient to identify the characteristic dynamical timescale with the HBI/MTI growth rate ($e$-folding rate), $\omega^2_{dyn}\equiv g\frac{\partial\ln T}{\partial z}$. Eqn. (\ref{eqn:HBI_MTIcondition}) then simplifies, in a purely vertical magnetic field, to
	\be \label{eqn:UnsimplifiedStability}
	- \frac{k_\bot^2}{k^2} g \frac{\partial \ln T}{\partial z} + \frac{B^2}{4\pi \rho}(\mbf{k}\cdot\mbf{b})^2 > 0
	\ee
	
	This equation is essentially a quantification of the competition between the effects of a destabilizing temperature gradient and the stabilizing influence of magnetic tension.

\section{Numerical Setup}\label{sec:numMeth}

	For our simulations we use the ATHENA MHD code \cite[]{stoneetal2008} which uses an unsplit Godunov method utilizing constrained transport \cite[]{gs2005} to preserve the divergence free nature of a physical magnetic field. To prevent unphysical transport of heat against $\nabla T$ we add a monotonic flux limiter to our problem algorithm according to the scheme laid out in \cite{S&H2007}.

	Atmospheres unstable to the HBI are those with a positive temperature gradient. To achieve this condition and maintain positive entropy gradient we set up the initial equilibrium configuration with temperature, density, and pressure given by power laws of the form
\begin{subequations}
\begin{align}
	T(z)&=T_0\left(  1+\frac{z}{H_{T}}  \right) \\
	\rho(z)&=\rho_0\left(  1+\frac{z}{H_{T}}  \right)^{-2} \\
	P(z) &= P_0\left(  1+\frac{z}{H_{T}}  \right)^{-1},
\quad \end{align}
\end{subequations}
with $H_{T} \equiv (d \ln T/dz)^{-1} = 2$ the characteristic scale height of the atmosphere with sound speed $c_s \equiv \sqrt{\gamma P/\rho}\approx 1.3$. From hydrostatic equilibrium, with $g=1$, we have $T_0 = 2 \rho_0 = P_0 = 2\mu = 2$ and we choose an initial magnetic field oriented purely vertically ($\mbf{B}=B_0\hat{z}$). 

	This equilibrium configuration has a conductive heat flux that is divergence free,
	\be
	\frac{d}{dz} \left( \chi_{\|} \frac{dT}{dz}  \right) =0,
	\ee
	since we choose a constant value for the parallel component of thermal conduction, $\chi_{\|}$. Aside from ensuring that the HBI will develop with the same growth rate everywhere in the anisotropically conducting part of the atmosphere, since our models do not account for radiative cooling and zero-divergence implies no heating, the atmosphere is initially in thermal equilibrium.
		
	Our simulations have dimensionless width$\times$height of 0.1$\times$0.3 and we limit anisotropic conduction and all our measurements of the gas dynamics to the central 0.1$\times$0.1 region. The remaining 0.1$\times$0.1 buffer zones above and below have isotropic conduction of the same magnitude as conduction in the active central zone. Early testing revealed the vertical reflective momentum boundary condition to cause suppression of the vertical growth by deflecting motion of the plasma before it naturally reached a non-linear state, but addition of buffer zones of this size alleviates the effects of the boundaries. 
	
	We choose momentum reflective boundary conditions for the upper and lower boundaries to provide mass conservation and pressure support. Periodic boundary conditions are chosen for the left and right boundaries. The temperature is held constant at the upper and lower boundaries where the magnetic field vector is set identically to that in the last active cell for each column. This magnetic boundary condition conserves total vertical magnetic flux.
	
	In order to resolve the linear development of the HBI cleanly, we applied single-mode perturbations to the equilibrium atmosphere in most of our simulations. For consistency, we tested the dependence of our results on the perturbation spectrum by running a set (the H2d128\_xPS series) of simulations with identical initial conditions but a power spectrum perturbation in momentum with $|v_k|^2\propto k^{-\alpha}$ of Kolmogorov type, $\alpha=5/3$. This produces a non-linear magnetic field topology approximately the same as a pure-mode perturbation where two wavelengths fit in each direction in the central square box. For all simulations exploring the $\beta$-resolution parameter space we start with a single pure-mode in both velocity and magnetic field of magnitude $k_x = k_z=4\pi/0.1$. 
			
	We performed 24 simulations to cover the $\beta$-resolution parameter space and use the designation H2d128\_5, for example, to label our 2-d HBI simulation with a resolution (covering the active region) of 128$\times$128 and an initial plasma $\beta$-parameter of $\beta=2\times10^5$. To connect our work with both linear theory and previous studies we chose five values of $log_{10}(\beta) \sim 3,5,7,9$ and $11$, and five grid resolutions $R\times3R$ where $R = [32,64,128,256, 512]$ \footnote{We chose not to use limited computing resources on the 25th simulation, H2d512\_3, since the HBI is stable for that value of $\beta$ and no dynamics is seen for any grid resolution.}. In addition to these 24 simulations we ran a power spectrum simulation of resolution 128$\times$384 for each $\beta$ and a 3-d simulation of resolution 128$\times$128$\times$384 of $\beta=2\times10^7$.

\section{Plasma Behavior as a Function of Resolution and $\beta$}\label{sec:Observations}

	In this section we show the results of our $\beta$-resolution parameter space study. We first describe the effect of varying $\beta$ (i.e., only changing magnetic field strength) on the non-linear saturation of the HBI. We then show the dependence of these effects, and evidence for general convergence of the physics of the HBI, with the grid resolution.
	
	As a precursor to our discussion of the full parameter space, we show in the first two frames in Figures \ref{fig:FourTimesTwoBeta}a and \ref{fig:FourTimesTwoBeta}b the temperature and magnetic field structure of our atmosphere under the pure-mode perturbation. These frames show the entire height of the atmosphere at times when the HBI is in the linear growth phase and just reaching the non-linear regime. Note that for both simulations of Fig. \ref{fig:FourTimesTwoBeta}a and \ref{fig:FourTimesTwoBeta}b, corresponding to values of $\beta=2\times10^7$ and $2\times10^9$ respectively, the linear growth phase appears identical, in agreement with the expectations of the linear theory.

	Now we focus attention on the development of the HBI into the nonlinear regime illustrated in the late-time frames of Figures \ref{fig:FourTimesTwoBeta}a and \ref{fig:FourTimesTwoBeta}b.

	\subsection{Plasma $\beta$}

	\begin{figure}   
	  \begin{center}
	      \includegraphics[width=0.5\textwidth]{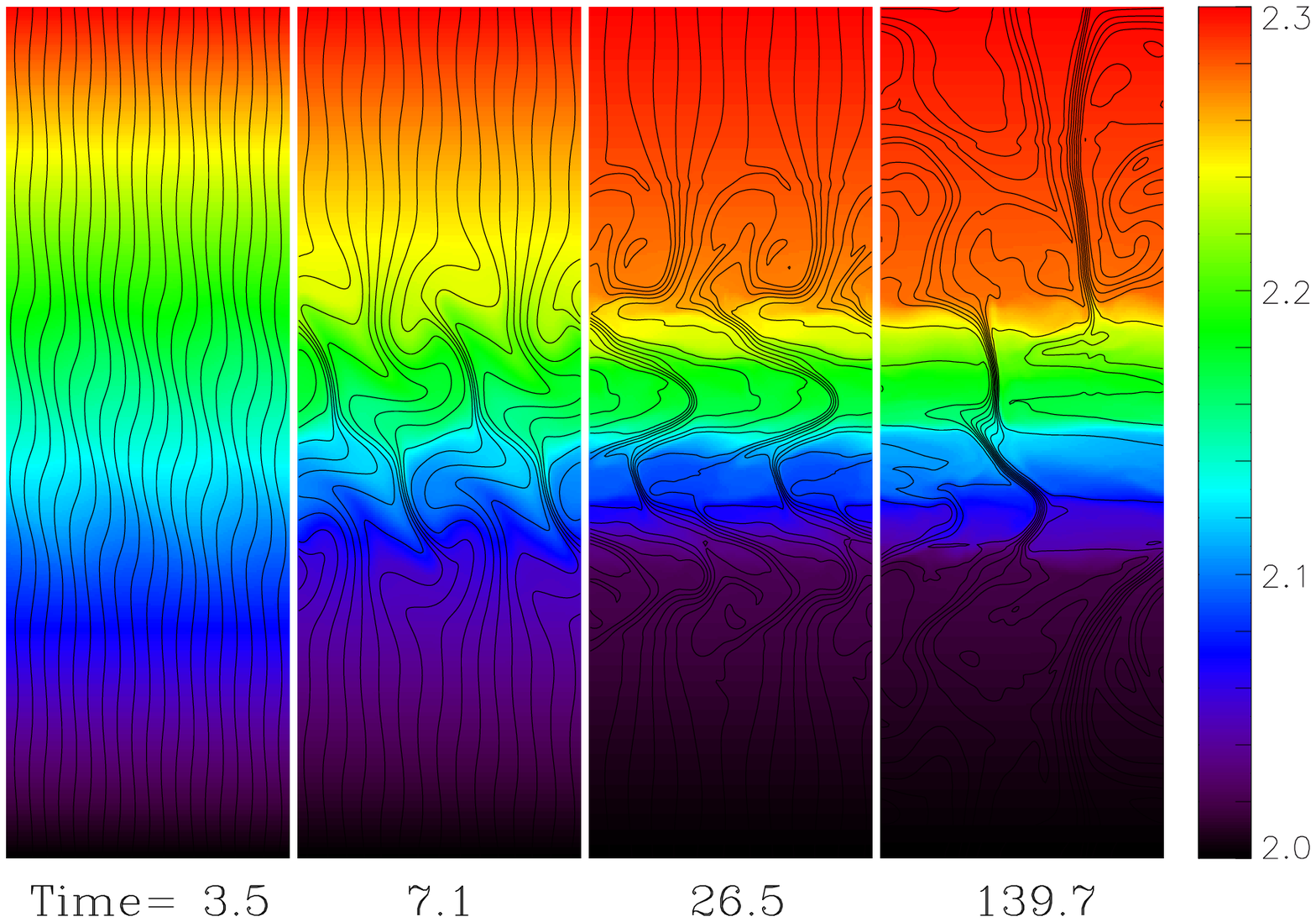} \\
	      \includegraphics[width=0.5\textwidth]{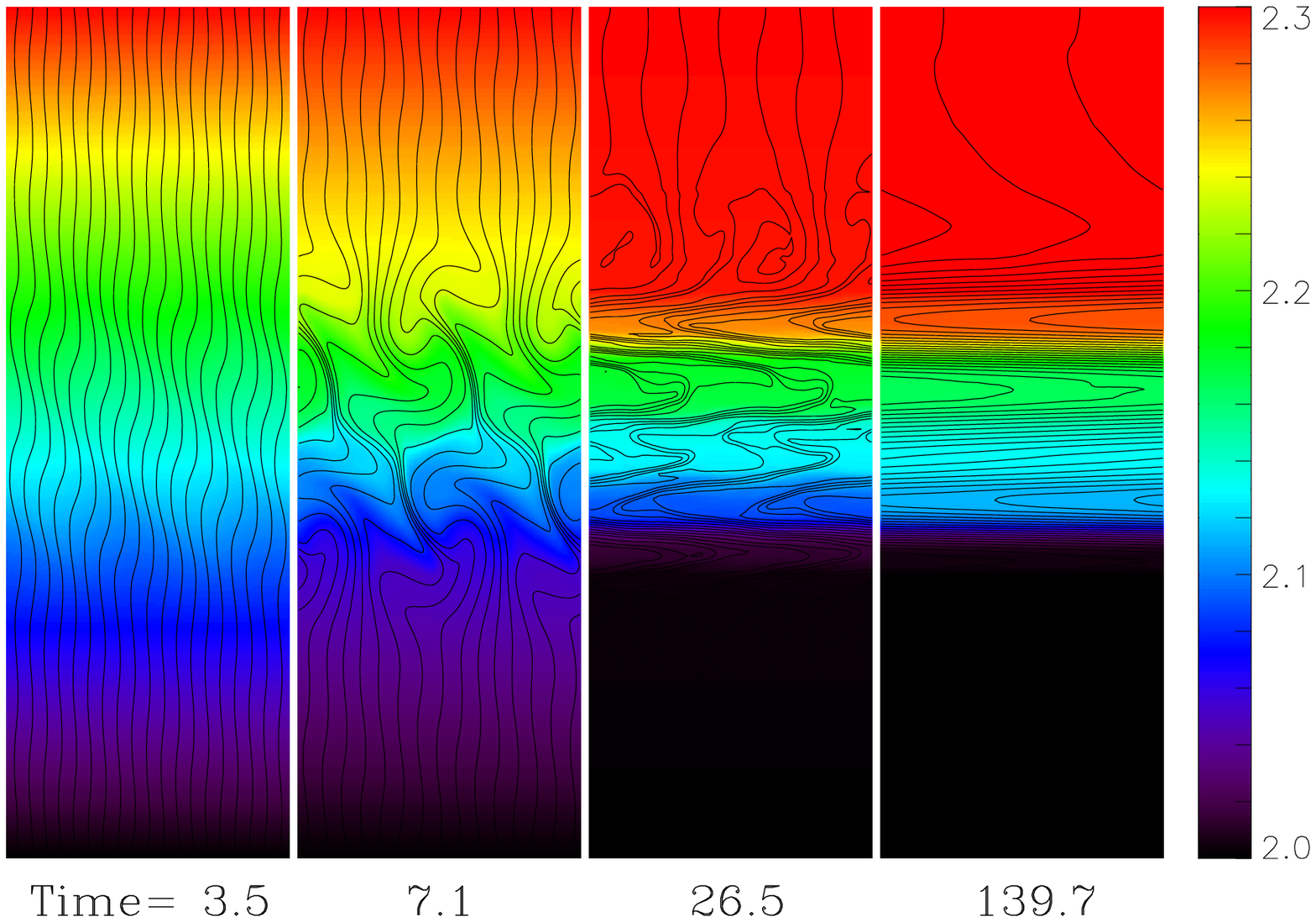} 
	       \caption[ ]{(a) H2d128\_7 at four times (in units of dynamical times, $\omega^{-1}_{dyn}$). The color bar on the right indicates temperature and magnetic field lines are overlaid. (b) H2d128\_9 at identical times as (a). Note the clear distinction between the two after many dynamical times, shown in the last frames.} 
	     \label{fig:FourTimesTwoBeta}
	  \end{center}
	\end{figure}

	Central to any study of the ICM is the plasma $\beta$-parameter. In the context of these simulations, consider the fully non-linear state of the plasma shown in the later frames in Figure \ref{fig:FourTimesTwoBeta}. The simulation with a moderate initial value of $\beta\approx10^7$ (H2d128\_7) contrasts qualitatively to that with $\beta\approx10^9$ (H2d128\_9).

	\begin{figure*}   
      		\includegraphics[width=1.0\textwidth]{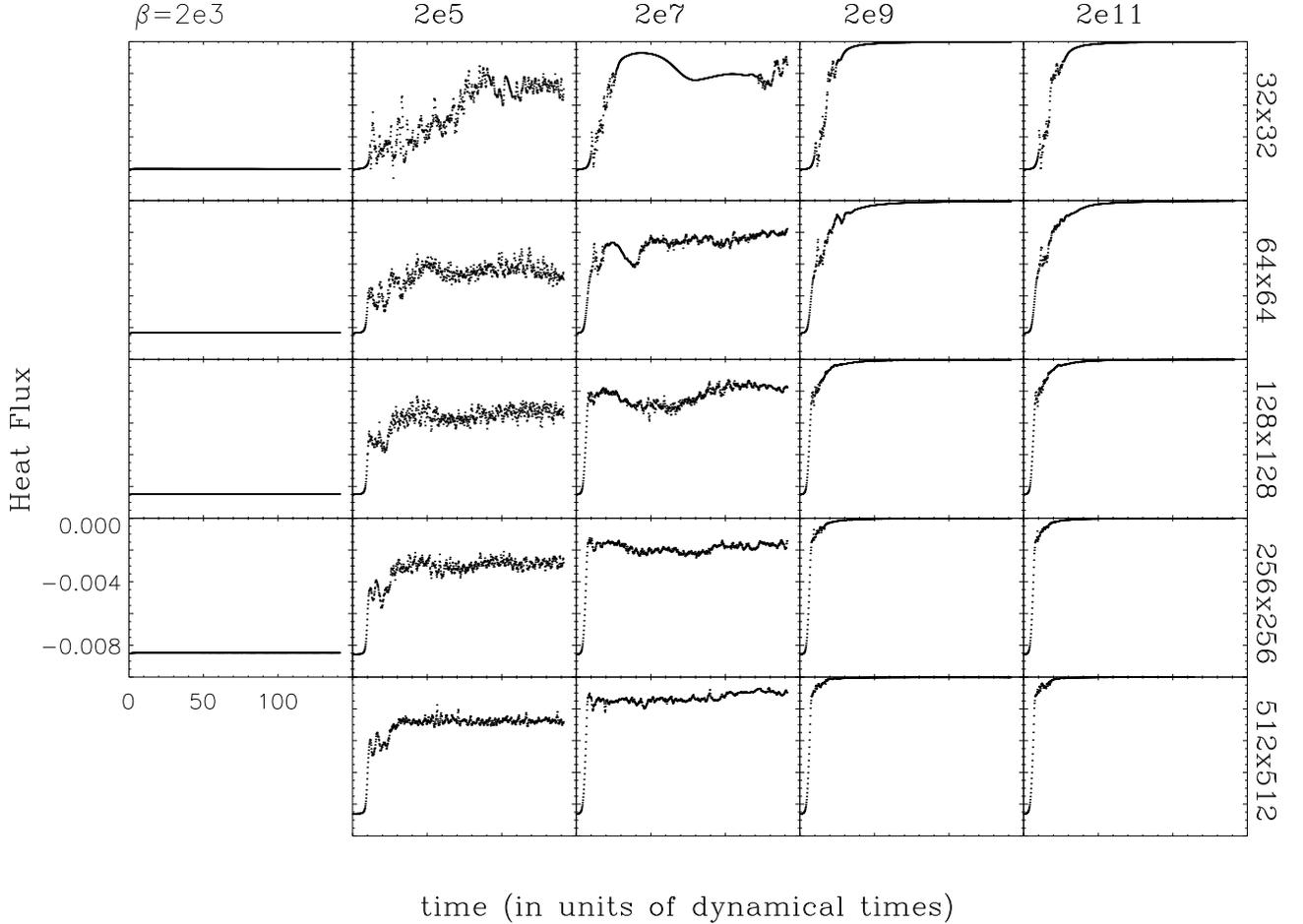}
      	 	\caption[ ]{Horizontally averaged vertical heat flux as a function of time for 24 simulations described in the text. Each panel in the plot has identically scaled axes.} 
    	 	\label{fig:VertCondAll}
	\end{figure*}
	
	The major qualitative difference is that magnetic flux bundles, or {\it filaments}, are present near the end of the linear phase in both simulations, but persist over many dynamical times only in the moderate $\beta$ case  ($2\times10^5$ and $2\times10^7$). These filaments have been seen before as the seeds of the horizontal bands of magnetic field lines which develop in high $\beta$ simulations. Horizontal motion, free from any restoring force \citep[see][for discussion]{2011MNRAS.413.1295M}, wraps the magnetic field lines thereby insulating temperature layers from one another and continuously growing in magnetic field strength. However, for the moderate values of $\beta$ these filaments persist against horizontal bulk motion and reach an entirely different saturated state. 
		
	The most important physical quantity characterizing the system, and an excellent discriminator of the different non-linear states, is the vertical heat flux (VHF),
	\[ F_z = -\int_S \chi \hat{b}(\hat{b}\cdot\nabla)T\cdot dS, \]
	where $S$ is a surface that cuts horizontally across the domain. Figure \ref{fig:VertCondAll} shows the horizontal average of the VHF across the center of the active domain ($z=1.5$) as a function of time for all simulations in the pure-mode study. The column containing those simulations stable to the HBI shows constant heat flux consistent with the magnetic field remaining vertical and conducting uniformly. The two intermediate $\beta$ columns show a nonzero heat flux in the saturated state. As shown in Fig. \ref{fig:FourTimesTwoBeta}, the HBI has shut off VHF in most of the plasma, but the filaments remain vertically oriented against the turbulent motion of the plasma and support a sustained VHF. The higher initial magnetic field strength results in a larger magnitude of sustained VHF, which we explain via the physical model for the filaments described in \textsection \ref{sec:FormDyn}. Finally, the high $\beta$ simulations lead to a non-linear state with zero VHF, duplicating the result of previous studies. Thus, one can identify the behavior in Fig. \ref{fig:FourTimesTwoBeta}a with the 2nd and 3rd columns of Figure \ref{fig:VertCondAll}, and Fig. \ref{fig:FourTimesTwoBeta}b with the last two.
	
	\begin{figure}   
      		\includegraphics[width=0.5\textwidth]{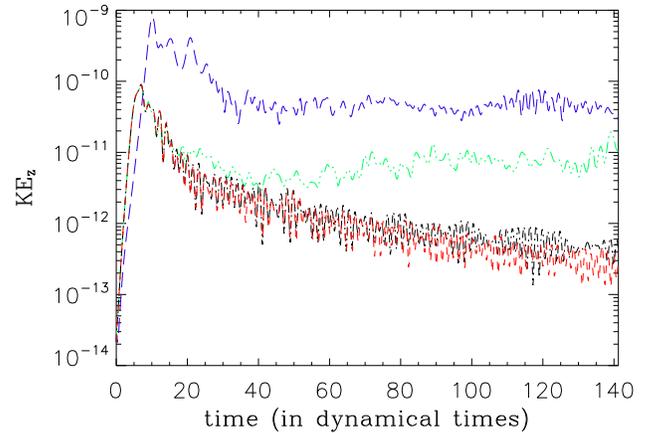}
      	 	\caption[ ]{The contribution to kinetic energy from vertical motion spatially averaged across the entire anisotropically conducting region is shown as a function of time for simulations H2d256\_x, x$\sim$[5 ({\it blue-long-dash}), 7 ({\it green-dash-dot-dot-dot}), 9 ({\it black-dash-dot}), 11 ({\it red-dash})]. } 
    	 	\label{fig:KEz}
	\end{figure}

	Another interesting quantity is kinetic energy. In order to study energy equipartition and turbulence in an inherently asymmetric atmosphere we break the kinetic energy scalar into the components contributed by motion in each direction, $KE_z\equiv\frac{1}{2}mv_z^2$ and $KE_x\equiv\frac{1}{2}mv_x^2$. Figure \ref{fig:KEz} shows how $\langle KE_z\rangle$, the vertical kinetic energy contribution spatially averaged across the entire active central region, varies with $\beta$. H2d256\_9 and H2d256\_11 show approximately linear damping of vertical motion, but H2d256\_5 and H2d256\_7 show an increase in vertical momentum with higher initial magnetic energy density. Both H2d256\_5 and H2d256\_7 saturate at a value orders of magnitude above the weak field cases, where the vertical momentum does not stop declining up to the end of our simulations. 
	
\begin{figure*}   
      		\includegraphics[width=1.0\textwidth]{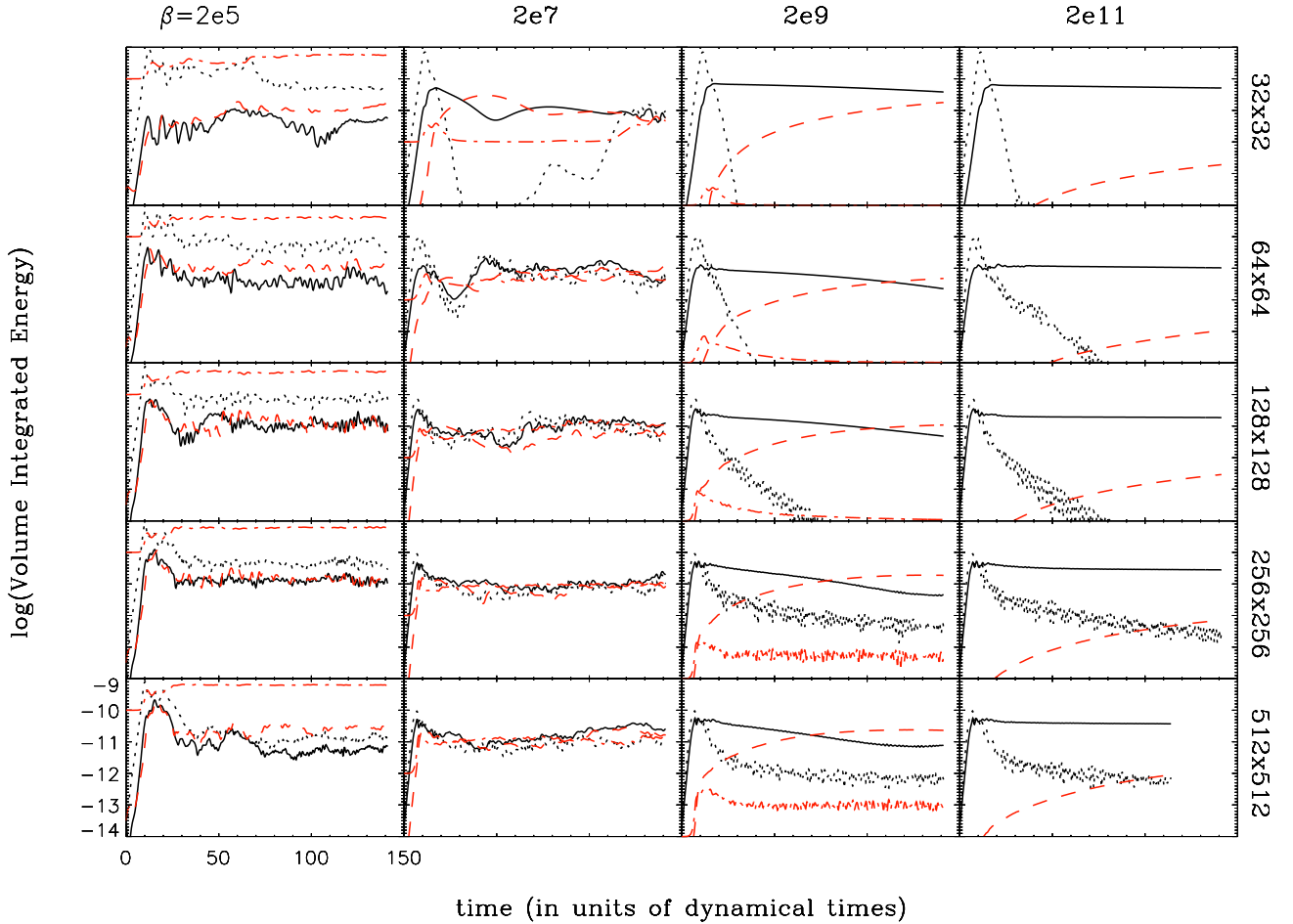}
      	 	\caption[ ]{Evolution of energy components, in our rationalized units, volume integrated over the active (central) region of the computational domain of dimensions 0.1x0.1x0.001. The last dimension is the arbitrary skin depth we choose for the compact third dimension. Only results of HBI unstable simulations are shown, namely, those for $\beta=2\times10^3$ are omitted. $KE_x$ kinetic energy ({\it black-line}), $KE_z$ kinetic energy ({\it black-dotted}), $B^2_x/8\pi$ magnetic energy ({\it red-dash}), and $B^2_z/8\pi$ magnetic energy ({\it red-dash-dot}). Each panel in the plot has identically scaled axes.}
    	 	\label{fig:Equipartition}
	\end{figure*}

	However, the kinetic energy is not the only energetic discriminant for the non-linear behavior of the plasma. Fig. \ref{fig:Equipartition} shows the vertical and horizontal contributions to the volume integrated magnetic and kinetic energy within the anisotropically conducting region. For $\beta=2\times10^9$ the kinetic energy completely dominates the dynamics of the plasma early on as horizontal motion wraps the field into layers. It is only at late times when the field wrapping has transferred enough energy into the magnetic field for the horizontal component $B^2_x/8\pi$ to overcome $KE_x$.
	
	The moderate $\beta$ columns in Fig. \ref{fig:Equipartition} show something very different. For sufficient resolution there is approximate equipartition between $B^2_x/8\pi \approx KE_x \approx KE_z$. In fact, for initial $\beta=2\times10^7$, even $B^2_z/8\pi$ reaches equipartition with the other energy components. In general, however, $B^2_z/8\pi$ is subject to a special constraint, namely total vertical magnetic flux through the simulation domain is conserved, and so does not necessarily come into equipartition.
	
	While the simulations start with four different initial field strengths spanning four orders of magnitude, the total of all the energy components (excluding $B^2_z/8\pi$) is the same for all simulations with resolution at least 64$\times$192, whether filaments are sustained or not. This is expected from the linear theory since the linear growth rate of pure-mode perturbations is independent of magnetic field strength (in the weak-field limit) and the linear growth entirely determines the amount of kinetic energy at the time of transition into the non-linear regime. From inspection of Fig. \ref{fig:Equipartition} it seems that the balance between volume integrated kinetic and magnetic energy soon after transition into the non-linear regime is an indicator of whether or not filaments are sustained. This makes sense as an excess in KE is necessary to overcome the tension within the filaments. Production of sustained filaments is intimately tied to the total kinetic energy, and thus the turbulent state of the plasma.

	\subsection{Grid Resolution}

\begin{figure}   
      		\includegraphics[width=0.47\textwidth]{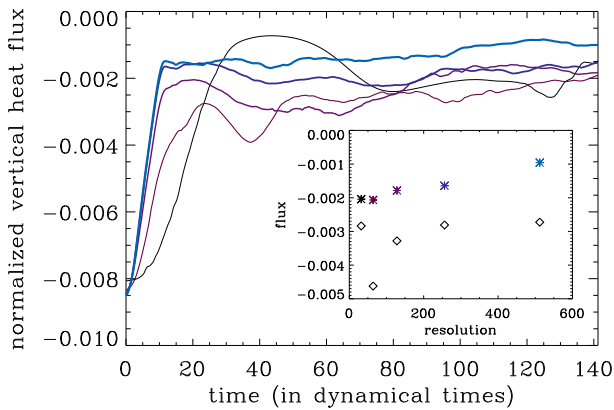}
      	 	\caption[ ]{Arbitrarily scaled, horizontally averaged vertical heat flux plotted in the $\beta\sim10^7$ column of Fig. \ref{fig:VertCondAll} versus time in dynamical units. Colors, in order of transition from black to blue, signify increasing resolution. {\it Inset:} Each data point represents the flux averaged over the last $18 \omega^{-1}_{dyn}$ for each grid resolution. Convergence is shown for simulations with initial $\beta\sim10^5$ (diamonds) but not $\beta\sim10^7$ (stars). }
    	 	\label{fig:VHFvsRes}
	\end{figure}

	Now we consider the more subtle effect of grid resolution as well as its interplay with $\beta$. The amplitude of VHF in the two intermediate $\beta$ columns of Fig \ref{fig:VertCondAll} decreases with increasing resolution. This is illustrated in Fig. \ref{fig:VHFvsRes} where the conduction rate vs. time for the $\beta=2\times10^7$ column at all five grid resolutions is smoothed and plotted in superposition. Inset is the time-averaged value of conduction during the last 18 dynamical periods for each resolution for $\beta=2\times10^5$ and $2\times10^7$. One can see that for $\beta\sim10^5$ there is negligible change in heat flux for the last doubling of resolution. However, for $\beta\sim10^7$ there is no such sign of convergence. While the data are too sparsely sampled in resolution to conclusively make this distinction, it is consistent with the general physical model of the filaments we propose in the next section. 
	
	Dependence on grid resolution is also evident in Fig. \ref{fig:Equipartition}. Firstly, for moderate $\beta$, increasing resolution produces less dispersion in the values of the four volume integrated energy components. Secondly, for high $\beta$ simulations there is an increase in the vertical energy components with increasing resolution. This second effect could be due to the resolution of small scale turbulence. 
	
	For high $\beta$ simulations it also takes less time with increasing resolution for $KE_x$ and $B^2_x/8\pi$ to switch dominance. This is consistent with a lower level of numerically driven magnetic reconnection in the higher resolution simulations and the regions of highest magnetic field curvature. Finally, in the $\beta=2\times10^5$ column, increasing resolution results in a lower saturated value of $KE_z$. In these simulations a significant fraction of the vertical motion is bulk flow along the filament, similar to what was seen in the filaments of K2012. We do not elaborate on the flow in this study, but it suffices here to simply say the width of filaments is observed to decrease with increasing resolution. Thus, the volume integrated $KE_z$ as well as the resolution dependence of VHF can be explained by a dependence of filament width on resolution. In the next section we present a physical model for the filaments and explain their dependence on the $\beta$-resolution parameter space.

\medskip
\medskip

\section{Dynamics and Persistence of Filaments}\label{sec:FormDyn}

	Why do filaments form and what sets their field strength and width? We hypothesize that filaments are the result of needing to pass a (conserved) vertical magnetic flux through a plasma subject to an instability which attempts to remove vertical field. We suggest that the filaments collapse until their internal field strength renders them HBI stable, with further collapse likely prevented by the diffusive process of turbulence. 

        Without viscosity, a plasma governed by ideal MHD will be driven by turbulent motions toward equipartition of the total kinetic and magnetic energies via field amplification. Therefore, after the filaments are produced, if the kinetic energy density is sufficiently higher than the magnetic energy density, the large scale bulk motions of the plasma (preferentially horizontal since the atmosphere is classically Schwarzschild stable) will wrap up the magnetic field into horizontal layers. If, on the other hand, the magnetic field dominates overall, most magnetic flux is vertical and the filaments are sustained. While the detailed dynamics of filament stability against the turbulent motions of the plasma could be framed in terms of horizontal velocity shear and height of the filaments, the complex magnetic topology and boundary limitations of the simulations suggest more insight may be gleaned from an understanding of filament persistence in terms of globally averaged quantities. Therefore, our model presents a global energetics understanding of filament persistence, and a separate model for filament width and field strength in terms of the HBI. This separation is natural since perturbations of wavelength greater than about twice the width of the filament are better understood geometrically in terms of bending the flux tube than in terms of the linear HBI. 

\begin{figure}   
      		\includegraphics[width=0.5\textwidth]{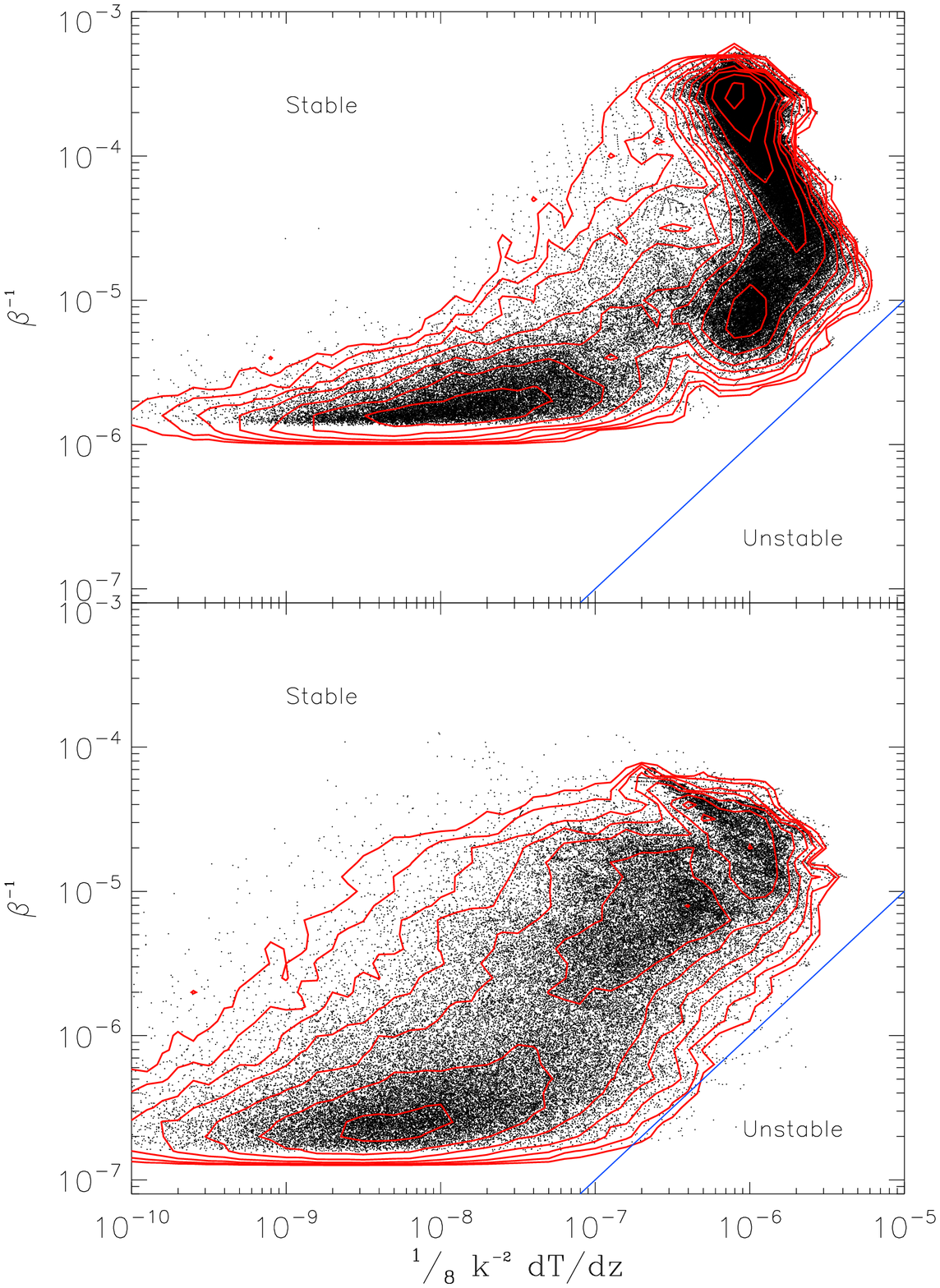}
		
      	 	\caption[ ]{ H2d512\_5 (upper) and H2d512\_7 (lower) filaments. Stability of the filaments is demonstrated by plotting the three stability diagnostics of Eqn. (\ref{eqn:SimplifiedStabilityRelation}) measured for each filament. Phase space below the solid line is HBI unstable, above is stable. The data points are evenly temporally sampled from the last $35.4 \omega^{-1}_{dyn}$. Those plotted are a subset of all extracted, for visual clarity, but the contours represent all the data and enclose $2^{[4,5,...17]}$ points.}
    	 	\label{fig:FilamentPhaseSpace} 
	\end{figure}

        A direct prediction of our model is that simulations should show that filaments lie on the stable side of the threshold between stable and unstable regions of phase space. Fig. \ref{fig:FilamentPhaseSpace} shows the result of such a measurement, obtained via the method we now describe.
	
	First consider the size of the filaments. Our hypothesized model predicts filaments will have a width corresponding to a wavenumber which just satisfies HBI stability, i.e., the filaments are stabilized when $\omega^2_A>\omega^2_{dyn}$ for a region inside the filaments of size equivalent to the scale of a random perturbation. 
	 
 	\begin{figure}   
      		\includegraphics[width=0.5\textwidth]{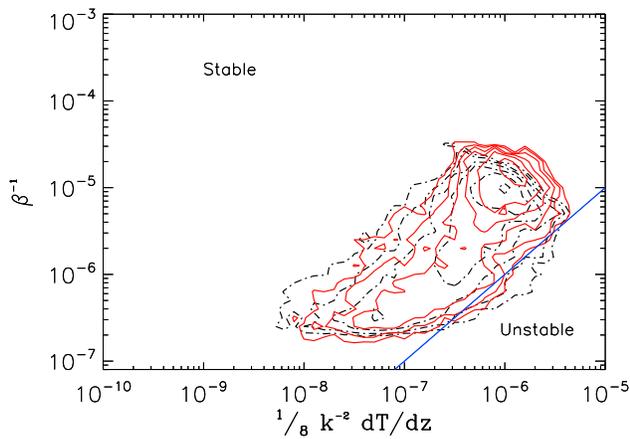}		
      	 	\caption[ ]{ Black contours (dash-dot) represent filaments from H2d128\_7PS; red (solid) represent those from H2d128\_7. Contour levels are set the same as in Fig. \ref{fig:FilamentPhaseSpace}. }
    	 	\label{fig:PurePScompare}
	\end{figure}
	
	In our analysis for Fig. \ref{fig:FilamentPhaseSpace}, we must be able to define and chracterize filaments in an automated manner. We consider a filament to be located wherever a series of horizontally adjacent grid cells satisfy both $|\hat{b}_z| > 0.7$ and $\beta_{cell} < 0.1 \langle \sqrt{\beta} \rangle^2 $, a weighted average of $\beta$ over the active domain. The values of magnetic energy (for which we can use $\beta^{-1}$ as a proxy since the average thermal pressure varies at most by 2$\%$), the vertical component of the temperature gradient, $\frac{dT}{dz}|_z$, and the filament width, which we identify with wavenumber $k_w$, are extracted for each such collection of cells for all but the rows of our anisotropically conducting active region directly adjacent to the buffer regions. Each set of values produces a point in Fig. \ref{fig:FilamentPhaseSpace}. In the plot the $k_w$ value used is calculated assuming a wavelength set at twice the measured width of the filament. The values of vertical temperature gradient within the filament and $\beta^{-1}$ are unweighted horizontal averages across the filament-defining cells. 
	
	 If we assume $b_x$ is close to zero, which is generally observed to be the case for sustained filaments, then it follows that $K =-k_\bot^2$. If we additionally assume $k_\bot\approx k_\|$ for an average perturbation, then $2k^2_\|\approx k^2$, and Eqn. (\ref{eqn:UnsimplifiedStability}) simplifies to
	 \be \label{eqn:FullStabilityRelation}
	 	-\frac{1}{2}\frac{g}{T^2}\frac{dT}{dz} + \frac{1}{\beta}k^2 > 0.
	 \ee
	 This provides a general relation giving the boundary of HBI stability in terms of our three diagnostic measures of the filaments. In order to facilitate comparison of the relation with the data in Fig. \ref{fig:FilamentPhaseSpace} we further simplify by setting $T=2$, in our local simulations where temperature does not significantly spatially vary. Also, since the width of the filament is the scale at which it is stabilized to random perturbations under our model, we identify $k$ as the width $k_w$. We then obtain
	 \be \label{eqn:SimplifiedStabilityRelation}
	 	\beta^{-1} \sim \frac{1}{8} \frac{dT}{dz} k_w^{-2}.
	 \ee 
	
	In Fig. \ref{fig:FilamentPhaseSpace} data points created in the method just described are extracted from the active region for the last $\sim35$ $\omega_{dyn}^{-1}$ of the simulation at 200 evenly spaced times. The magnetic filaments are clearly subject to the boundary provided by the stability relation, thus supporting the theory that the filaments are islands of HBI stability within an otherwise HBI susceptible plasma. Specifically, consider how the filaments in H2d512\_7 and H2d512\_5 are scattered along the stable side of the curve but generally follow it. While there is a significant vertical spread in $\beta^{-1}$ above the line, consider that initial values in these simulations were $\beta^{-1}=5\times10^{-8}$ and $5\times10^{-6}$ and the final spread in $\beta^{-1}$ above the intra-filament value is consistent with the theoretical boundary over nearly three orders of magnitude. Note that some filament values of $\beta^{-1}$ fall below the initial conditions value. This is consistent with our definition of a filament having significant internal magnetic energy density over the spatial average, and most space in the atmosphere has a very weak field in the saturated state.
	
	Figure \ref{fig:PurePScompare} compares filaments from H2d128\_7 and H2d128\_7PS showing that they cover roughly the same region in this space. This suggests the pure-mode perturbation recovers most of the physics of a more multi-mode initial condition, and both lie at weaker field strengths than the simulations with higher resolution, suggesting the filaments in the lower resolution simulation are prevented from shrinking to obtain their natural peak field strength. This picture is also consistent with the H2d128\_7 data points residing at the higher values of $k_w^{-2}$ on the plot. 
		
	This self-consistent model for the internal stability of the filaments is sufficient to explain the width of the filaments and their dependence on magnetic field strength and resolution, as we've shown. However, it also predicts the viability of magnetic filaments in a plasma of arbitrarily weak magnetic field so long as there is sufficient resolution for the filament to shrink until it becomes internally HBI stable. This would predict that the boundary between our qualitatively different non-linear states would shift to higher $\beta$ as resolution is increased in Fig. \ref{fig:VHFvsRes}. While it is likely we inadequately resolve the $\beta$-resolution phase space to see such a trend to begin with, the true dissolution of this discrepancy is evident upon closer inspection of the global energetics in our simulations.
	
	Recall that both the two intermediate and two high $\beta$ columns of Fig. \ref{fig:VertCondAll} are well within the weak-field regime (See \textsection\ref{sec:MHD}). The growth rate then only depends on the pure-mode perturbation wavelength, which is identical for all simulations in our phase-space suite. Transition to non-linearity happens at the same time in all cases, and the exponential growth in momentum during the linear phase is driven by a conversion of heat into kinetic energy. These facts together imply that the total kinetic energy imprinted on the plasma at the end of the linear regime is the integrated net heat-flux passing into the active region during that time. For these reasons, and evident in Fig. \ref{fig:Equipartition}, the peak in kinetic energy at the end of linear growth is identical in all the simulations. Those with initial $\beta\sim10^5$ do have a slightly higher transition value, but that is consistent with the additional magnetic tension maintaining a pure-mode magnetic topology for slightly longer, therefore increasing the time over which the energy conversion takes place. Clearly the peaks in KE components for that column occur at later times than in the other columns.  
	
	In the second part of our model we suggest that the global persistence of filaments once they've formed depends on how the imprinted kinetic energy just described compares to the initial total magnetic energy. We argue that if approximately twice the magnetic energy density is greater than the kinetic energy density imprinted at the transition to non-linearity, then sustained vertical filaments of the form we see will be present. This is effectively a statement that either the magnetic field or the turbulent motion dominates and one or the other wins out early in our simulations, leading to a distinct division in parameter space. The `twice' comes from allowing an x-component of the filaments to grow to equal magnitude with the vertical z-component. Beyond that, the filament would be more horizontal than vertical, significantly quench vertical heat flux, and effectively be stretched out beyond our definition of a filament. 
	
	This scale-invariant method for understanding filament stability is particularly suited to a local study such as ours. Of course, it suggests that a sustained source of turbulence may significantly alter the non-linear state of the plasma. We leave a discussion of this and other implications for the last section.

\section{3-D Simulation}\label{sec:3D}
	
	Some studies suggest \cite[]{2012MNRAS.422..704P} that the filaments formed in 2-d simulations are a result of field lines not being able to slip past one another, that is, a prevention of flux-interchange modes in the plasma. K2012 comments on this in their H3dBrag simulation, where the filaments become more reoriented (horizontally) in the lower part of the atmosphere where the effects of Braginskii viscosity are less pronounced. However, this region of H3dBrag is very close to the simulation boundary, and the effect of the boundary on the detailed field structure is unclear.
	
	Our model explaining the presence of the filaments, presented in the previous section, suggests this existence is rooted in the physics of the HBI and not a feature of dimensionality. K2012 does not provide a complimentary 3-d semi-global simulation of the HBI without viscosity and so in extension of that work, and to confirm the robustness of our results, we performed a 3-d simulation of 128x128x384 resolution and moderate field strength, H3d128\_7PS. The simulation was in every way identical to H2d128\_7PS save for the added dimension.
	
\begin{figure}   
 		  \begin{center}
      			\includegraphics[width=0.5\textwidth]{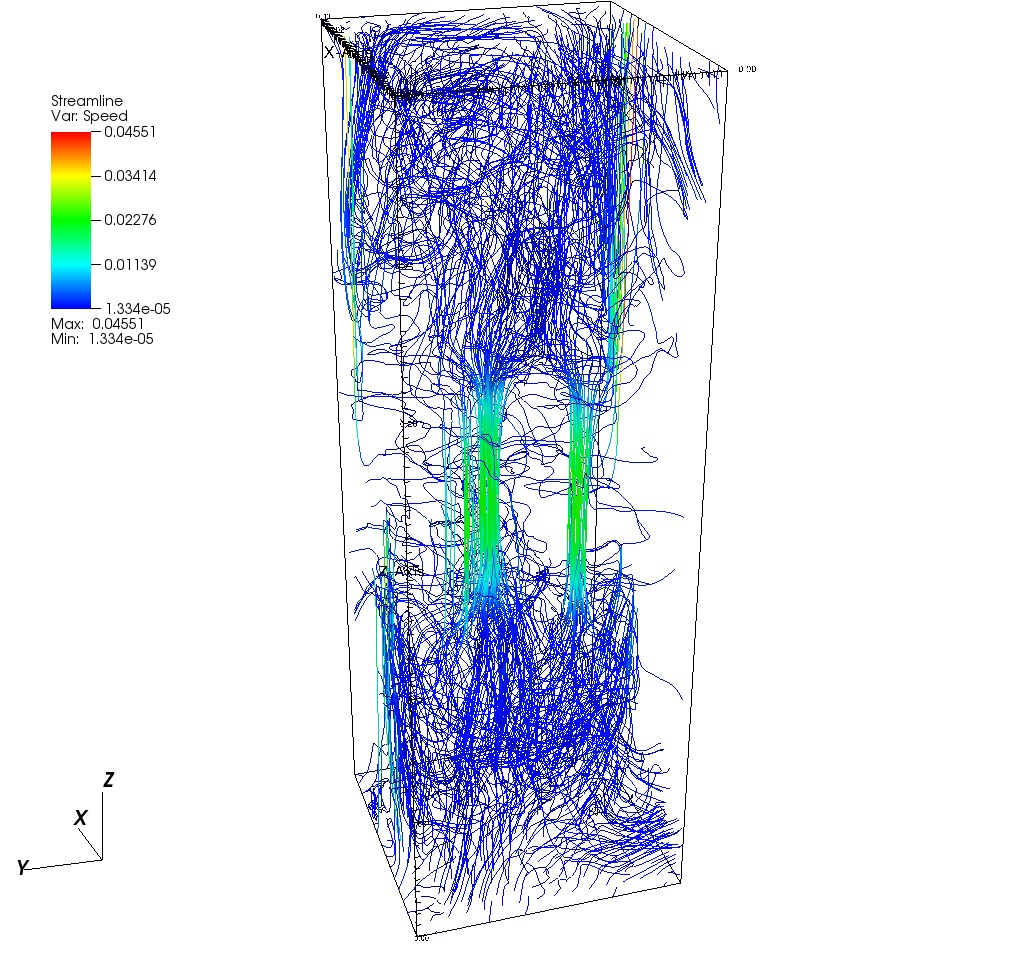}
      	 		\caption[ ]{Shown is a frame from the saturated state of H3d128\_7PS. Color value indicates magnetic field strength and field lines are integrated from grids of points at the top and bottom of the domain. Many of the field lines thread a filament near the center of the anisotropically conducting region.} 
    	 		\label{fig:H3d128_7PSimage}
  	  	  \end{center}
		\end{figure}

	Figure  \ref{fig:H3d128_7PSimage} shows the magnetic structure of the non-linear state of H3d128\_7PS at late times, with coherent filaments of a qualitatively similar sort to those found in the 2-d analog. Just as is pointed out in K2012, these filaments contain more magnetic flux because it is easier to collect with the additional dimension during linear growth. 
	
\begin{figure}   
 		  \begin{center}
      			\includegraphics[width=0.5\textwidth]{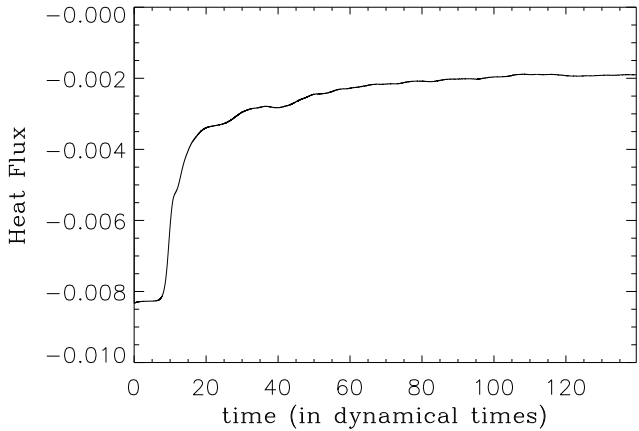} 
      	 		\caption[ ]{Vertical Heat Flux integrated over the horizontal plane slicing through the center of the active domain in H3d128\_7PS.} 
    	 		\label{fig:3DVHF}
  	  	  \end{center}
		\end{figure}

	The VHF measured in the same way as before is shown in Fig. \ref{fig:3DVHF}. H3d128\_7PS clearly runs long enough for the VHF to stabilize at a value of $\approx25\%$ the initial value, just as before. This value is no coincidence. The addition of a third dimension does not change the internally stabilizing $\beta$ of the filaments, and so the ratio of conducting surface area to non-conducting is still set by the initial plasma $\beta$ and total vertical flux conservation, assuming that most vertical flux is bundled into the filaments.
	
\section{Discussion and Conclusions}\label{sec:Discussion}

	In this paper, we present a systematic investigation of the local non-linear behavior of the HBI as a function of initial magnetic field strength and numerical resolution. We identify a previously unrecognized regime of behavior at intermediate field strengths in which magnetic filaments are formed that can resist the horizontal field line wrapping seen in weak-field simulations \citep{P&Q2008,2011MNRAS.413.1295M} and lead to sustained conductive heat flux. We show that the filaments themselves are HBI-stable and provide the main conduit by which (conserved) vertical magnetic flux is passed through the domain. In addition to this model for internal stabilization, we describe how global energetics determine the ultimate qualitative state of the non-linear regime, that is, whether or not filaments persist and there is a resulting significant vertical heat flux. 
		
	The robust formation of filamentary structures is tantalizing since we see this type of structure in the cooling cores of galaxy clusters, namely in the form of H$\alpha$ filaments. For instance, HST observations of the Perseus cluster, which has been studied closely for some time \citep{lynds1970}, reveal complex braided filaments which can be as narrow as 50 pc but may be kiloparsecs in length \citep{fabianetal2008}. More recent observations by \citet{mcdonaldetal2010, mcdonaldetal2011} have shown that essentially all cool-core clusters exhibit extended H$\alpha$ systems which appear to extend out to almost, but never beyond, the cooling radius of the X-ray emitting ICM. This connection between the cooling radius and the filamentary structure suggests they may condense out of the ICM as a result of thermal instability \citep{1965ApJ...142..531F}. 
	
	However, it is the {\it local} thermal instability which produces cold filaments in simulations, with anisotropic conduction primarily responsible for their filamentary structure. Without a heating source to offset the background global thermal instability, namely, when net cooling is purely a function of local density and temperature of the gas, buoyancy effects stabilize the plasma to the local thermal instability \citep{1989ApJ...341..611B}. K2012 shows that for an atmosphere with intermediate magnetic field strength and anisotropic viscosity, viscous heating during formation of the filaments and a sustained vertical heat flux can significantly delay the cooling catastrophe resulting from global thermal instability. This allows local thermal instability to develop and cold filaments to naturally form. Our study supports K2012, firmly establishing the robust formation of filaments over a range in magnetic field strength relevant to cool-core clusters, and exploring in detail the magnitude of vertical heat flux sustained by the atmosphere and its dependence on local magnetic field strength and numerical resolution. 
	
	Though our work has not yet shown a definitive connection to the observed H$\alpha$ filaments, in part because we do not include radiative cooling in our simulations, our findings strongly motivate further work in developing physically realistic global simulations of cluster cores, where the ICM is likely HBI unstable. Since filaments provide a mechanism for transporting heat (and possibly cold gas) radially in these atmospheres, an understanding of the magnetic topology from the local to the global scale and how these scales connect will be instrumental in understanding the role of conduction in regulating global thermal stability. Also instrumental is the interplay between the filaments and astrophysical turbulence, since turbulence can ultimately destroy otherwise stable filaments. Our discussion of the presence of filaments as a result of total energy balance provides some insight into the level of turbulence allowed when filaments are present, but more testing, especially on the semi-global and global scales is required.  
	
	We find, however, that to fully resolve the physics of the non-linear HBI in the local regime requires a high grid resolution. Thus, feasible global simulations of the ICM may require semi-analytic prescriptions for the dynamics of anisotropic conduction as well as anisotropic viscosity. This study may aid in production of those models since filaments have a key role in the thermodynamics of anisotropically conducting plasma.
	
	Another motivation for this work was to find, if any exist, robust measures for convergence with resolution. Given the intimate connection between the topology of the filaments and more global quantities such as heat conduction, it is not surprising that we find that adequate resolution of the dynamics of the filaments is needed in order to reach convergence in the plasma diagnostics such as VHF and equipartition. The most stringent measure we find is to measure the position of the filaments in HBI stability space presented in Fig. \ref{fig:FilamentPhaseSpace}. However, a much simpler measure that also performs well is to see how much the VHF changes with increasing resolution. 
	
	Should the HBI be driving dynamics in relatively large regions of real clusters, numerical studies including this one suggest that it may have a key role in cluster thermodynamics. While there is clearly an interaction between AGN and the ICM (e.g. radio lobes, etc.), it is not known exactly how this feedback may couple effectively to the thermodynamics. This work and K2012 suggest there may still be a potential for conductive solutions to the cooling flow problem. Certainly at this point, the field is still in the exploration stage of the fundamental physical processes which work together to govern the ICM.

\medskip

M.J.A., C.S.R., and T.B. acknowledge support from the NSF under ARRA grant AST-0908212. Support for T.B. was in part provided by the National Aeronautics and Space Administration through Einstein Postdoctoral Fellowship Award Number PF9-00061 issued by the Chandra X-ray Observatory Center, which is operated by the Smithsonian Astrophysical Observatory for and on behalf of the National Aeronautics and Space Administration under contract NAS8-03060. HPC resources were provided by the Texas Advanced Computing Center (TACC) (Ranger), as well as the UMD OIT (Deepthought) and Dept. of Astronomy (Yorp) clusters. The authors thank Matthew Kunz and Steve Balbus for useful conversations.

\end{document}